# Design of an electron gun for terahertz radiation source[†]

LI Ji(李骥)[1][*], PEI Yuan-ji(裴元吉)[1], HU Tong-ning(胡桐宁)[2], CHEN Qu-shan(陈曲珊)[2], FENG Guang-yao(冯光耀)[3], Shang Lei (尚雷)[1], Li Cheng-long(李成龙)[1]

[1] *National Synchrotron Radiation Laboratory, University of Science and Technology of China, Hefei 230029, China;*
[2] *State Key Laboratory of Advanced Electromagnetic Engineering and Technology, Huazhong University of Science and Technology, Wuhan 430074, China;*
[3] *Deutsches Elektronen-Synchrotron, Hamburg 22607, Germany*

**Abstract:** With the aim to obtain short-pulse bunches with high peak current for a terahertz radiation source, an EC-ITC (External-Cathode Independently Tunable Cells) RF gun was employed. As the external injecting electron source of the ITC RF gun, a gridded DC gun plays a key role, the performance of which determines the beam quality in the injector and transport line. In order to make the beam well compressed in the ITC RF gun, the energy of the electrons acquired from the gridded DC gun should be 15 KeV at most. A proper structure of the gridded gun is shown to overcome the strong space-charge force on the cathode, which is able to generate 6 μs beam with 4.5A current successfully.
**Keyword:** gridded DC gun, RF gun, EC-ITC,

## 1 Introduction

Terahertz wave, which is electromagnetic radiation between far infrared wave and microwave, has very important academic and application value. Compared to other Terahertz sources, FEL is the best way to get maximum power output. In this paper, a high-quality electron gun is studied for a compact FEL Terahertz radiation source.

Although photocathode RF guns have widely developed as electron sources for FELs, thermionic RF guns are expected to have potential to generate high brightness and short-pulse electron beams. The thermionic RF gun with two independently tunable cells (ITC) has been developed in National Synchrotron Radiation Laboratory for years, of which the cells are power fed independently. The external injecting ITC RF gun can generate beam bunches of superior characteristics by setting appropriate feeding powers and phases of the two cells instead of using alpha-magnet. And the external injecting structure can increase beam current and decrease energy spread and the negative effect of back bombardment to cathode.

In order to obtain beam pulse with sharp rising edge and falling edge, a 15 KeV grid-control DC gun was designed. The electrodes were properly distributed to generator 4.5 A beam current.

## 2 The external-cathode structure

As the power into the cells are fed independently, the ITC RF gun can capture the beam and compress the bunches well by setting the appropriate feeding power and launch phases separately, instead of using α-magnet or complicated laser drive system. In addition, the external-cathode structure we use can increase the captured beam current and decrease the energy spread than the common structure with cathode inside the cavity. And the negative effect of back bombardment to the cathode is nearly eliminated [4].

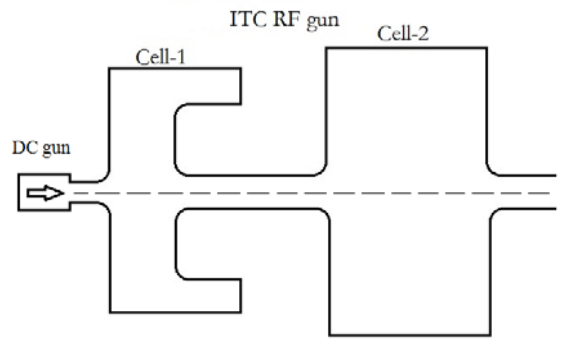

Figure 1: Layout of the external injecting ITC RF gun.

*Corresponding author (email:liji51@mail.ustc.edu.cn)
[†]Work supported by NSFC No. 10875116

## 3 The 15 keV gridded DC gun

### 3.1 Emitter

Considering the emission capability, the EIMAC Y646E cathode-grid assembly was chosen as the electron source. The stability and long life of this kind of cathode-grid assembly have been proved in testing and operation. The area of the emitting surface is 1 cm$^2$.

### 3.2 The double-anode gun model

In order to inject the 4.5 A beam into the ITC RF gun without hitting the wall of the cavity, the transverse beam size should be very small (<1.5mm). Thus, a gun of high perveance, high compression ratio and long shot distance should be designed. We have used a special geometry with intermediate electrode, which is also called double-anode structure, to satisfy the strict requirements.

Due to the strong longitudinal space-charge force of 4.5A beam exerted on the emission surface, the electric field strength should be large enough to extract the beam from the cathode-grid assembly. The beam from flat emission surface also need sufficient focusing force to get compressed transversely. Furthermore, the 15 keV/4.5 A beam is easy to diverge in drift space. To sum up, the Pierce-type diode structure is very hard to emit 15 keV/4.5 A beam into the ITC RF gun.

In order to enhance the electric field strength on the emission surface and the transverse focusing force, we add an intermediate electrode to accelerate the electrons to 35 keV, and then the electrons will be decelerated to 15 keV between intermediate electrode and the anode. This way the beam is focused twice in the gun. The focusing principle is shown in Fig. 2, in which the dotted lines represent the equipotential lines and the red arrowed lines mean the electric field lines.

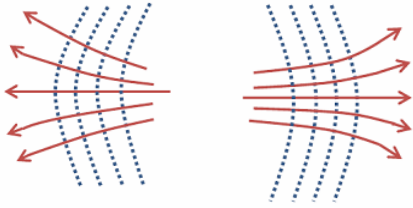

Figure 2: Focusing principle in acceleration (left) and deceleration (right).

The electric field distribution is calculated by Poisson code [6] and shown in Fig. 3, in which the red solid lines are the equipotential lines. The cathode-grid assembly is represented by emitter in this model, and the anode is a combination of real anode and a part of the first cell of the ITC RF gun.

The transverse electric field can be expressed as [7]:

$$E_r(r,z) \simeq -(\frac{r}{2})[\frac{\partial E_z(0,z)}{\partial z}] = -(\frac{r}{2})E_z^{'}. \quad (1)$$

where $E_z$ is the longitudinal electric field on the axis. As shown in Fig. 4, $E_z$ and $E_z^{'}$ the in the accelerating and decelerating process

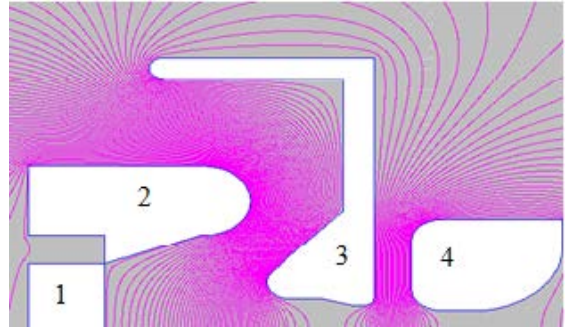

Figure 3: Model of 15 keV DC gun. From 1 to 4, the electrodes are emitter, focusing electrode, intermediate electrode and anode.

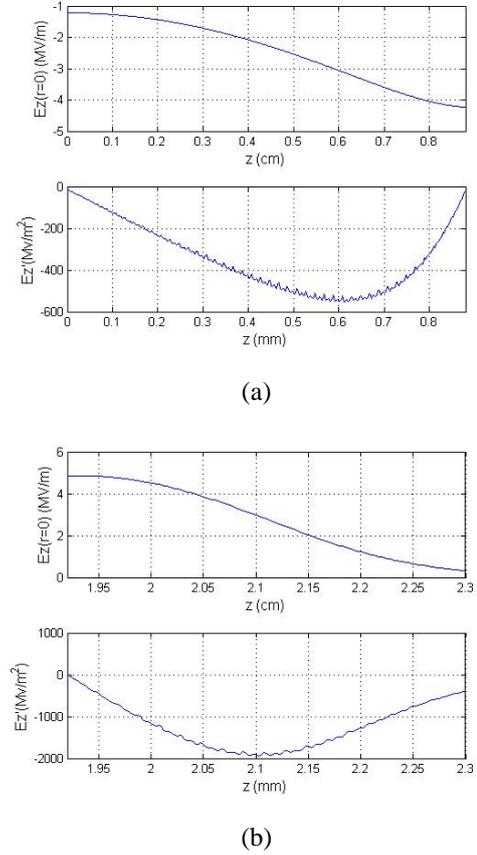

(a)

(b)

Figure 4: $E_z$ and $E_z^{'}$ on the axis in acceleration process (a) and deceleration process (b)

### 3.3 Simulations of the DC gun

The DC gun was modeled with OPERA-3D [8], of which the SCALA module analyses electrostatic fields taking into account the effects of space charge created by beams of charged particle. In this model, emitter and focusing electrode are connected with -15 kV and 20 kV power supplies separately. Considering the strong space-charge force of low-energy and high-density beam, the grounded anode is amounted together with a part of the first cell of ITC RF gun to decrease the distance in the anode hole. The initial condition is set as that 4.5 A beam with 150 eV extracted from the emission surface.

The simulation results in Fig. 5 shows that the 15 keV/4.5 A beam can be injected into the ITC RF gun sucessfully.The trajectory shows that the beam is foucsed at the entrance of the anode again to supress the divergence.The focusing strength is modilated through opitimize the shape of anode to avoid over focus and under focus. Briefly, the beam is shaped well in accelerating process and contained in the decelerating process.

The maximum electric field strength is 12.5 MV/m on the nose of intermediate electrode, and experiments show that breakdown threshold is 15 MV/m in DC case [9]. In operation, the emitter will be connected with 15 kV pulse power supply, which can reduce the possibility of breaking down.

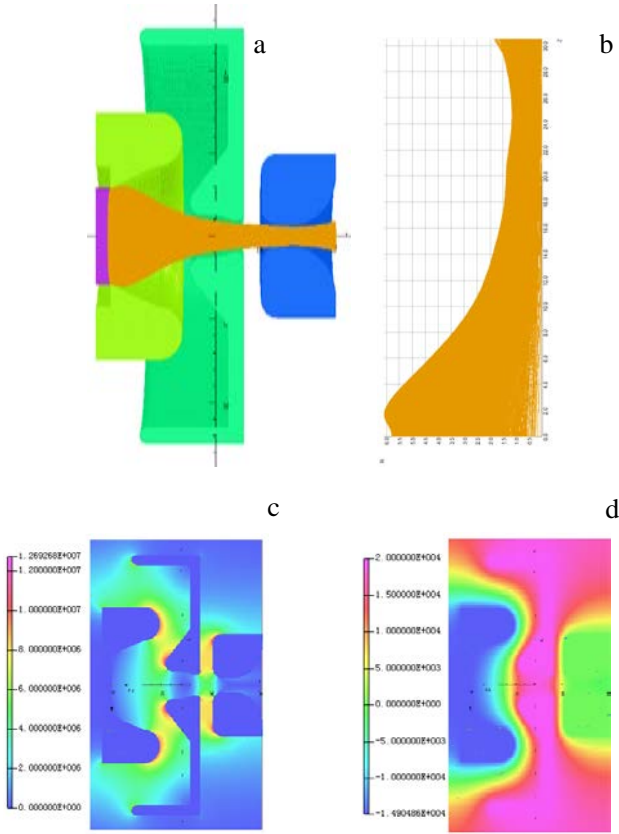

Figure 5: a. 3D model with trajectory b. Beam trajectory c. Electric field strength distribution d. Potential distribution

The simulation results are also testified by CST Particle Studio [10].The evaluation of electrons motion in Fig. 6 is almost the same with the trajectory from OPERA-3D.The normalized RMS emittance at the exit of the 15 keV gridded DC gun is 6.62 mm mrad.

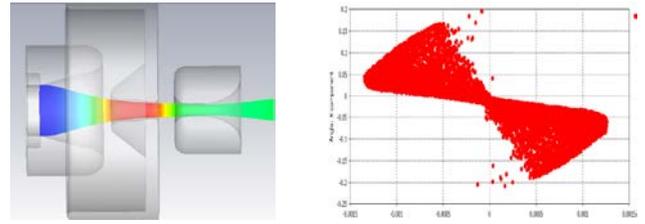

Figure 6: Beam trajectory in the 3D model (left) and phase space (right) from CST

### 3.4 The ITC RF gun

The beam from 15 kV DC gun is captured and bunched in the first cell of ITC, and then the beam bunches are compressed and accelerated to 2.6 MeV in the second cell. The two cells are power fed independently and achieve resonance at 2856 MHz .The magnitude of electric field strength along the axis in the cavities is calculated by Superfish code [6] in the Fig. 7, while the expected properties of the output bunch are listed in Tab. 1.

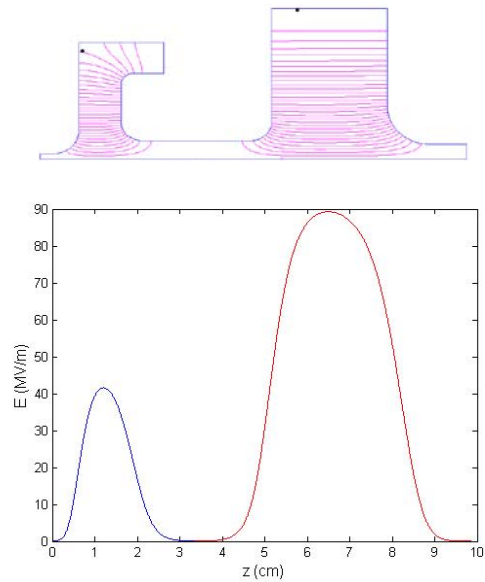

Figure 7: Superfish model of the ITC RF gun and magnitude of electric field on axis

Table.1 Expected properties of Output bunch of ITC RF gun

| | |
|---|---|
| Transverse emittance | < 10 mm mrad |
| Energy spread (FWHM) | < 0.50% |
| Micro-pulse length (FWHM) | 1~10 ps |
| Micro-pulse effective charge | 200 ~ 300 pC |

## 4 Particle-dynamics simulation

Dynamics computation from the DC gun to the ITC RF gun was executed with PARMELA code [11]. The performance of output is characterized with the parameters of effective part (the head) of the bunch, of which the length is ~10ps. The results of dynamics computation at the exit of the ITC RF gun are shown in Fig. 8 and Tab.1.

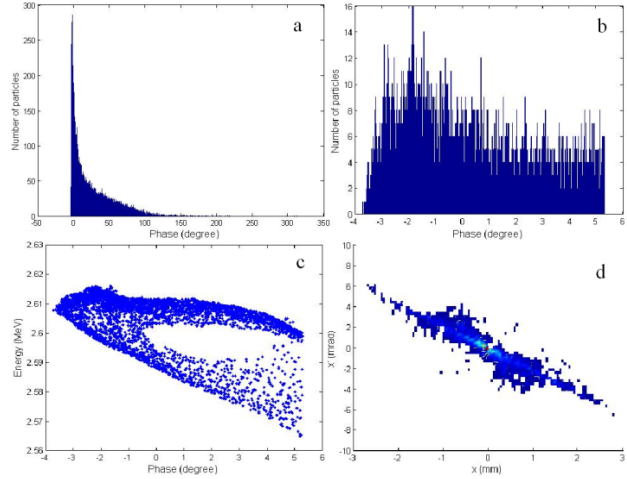

Figure 8: a. Phase spectrum (output) b. Phase spectrum (effective part) c. Energy vs. phase (effective part) d. Phase space (effective part)

Table.2 Properties of Output bunch

| | |
|---|---|
| Beam radius | 2.8 mm |
| Beam energy | 2.6 MeV |
| Transverse emittance | 9 mm mrad |
| Energy spread (FWHM) | 0.30% |
| Micro-pulse length (FWHM) | 3.8 ps |
| Micro-pulse effective charge | 210 pC |

## 5 Conclusion

A new-type gridded DC gun of high perveance, high compression ratio and long shot distance was designed for ITC RF gun. The simulation results show that the external-cathode RF gun can generate high-quality beam to satisfy the strict requirements of terahertz source. The effective bunch charge is over 200 pC with ~4 ps micro-pulse width (FWHM), while energy spread (FWHM) is ~0.3% and the transverse normalized RMS emittance is less than 10 mm mrad.